\documentclass[12pt]{article}
\usepackage{amsmath}
\usepackage{color}
\usepackage{amssymb}
\usepackage{bm}

\makeatletter\@addtoreset{equation}{section}
\makeatother
\allowdisplaybreaks[1]
\begin{document}
\begin{titlepage}

\begin{flushright}
\phantom{preprint no.}
\end{flushright}
\vspace{0.5cm}
\begin{center}
{\Large \bf
Teukolsky-like equations with various spins\\
\vspace{2mm}
in spherically symmetric spacetime
}
\lineskip .75em
\vskip0.5cm
{\large Ya Guo${}^{1}$, Hiroaki Nakajima${}^{2}$ and Wenbin Lin${}^{1,\,2,\,*}$}
\vskip 2.5em
${}^{1}$ {\normalsize\it School of Physical Science and Technology, Southwest Jiaotong University, \\ Chengdu, 610031, China\\
}
\vskip 1.0em
${}^{2}$ {\normalsize\it School of Mathematics and Physics, \\
University of South China, Hengyang, 421001, China\\}
\vskip 1.0em
${}^{*}$ {\normalsize\it Email: lwb@usc.edu.cn\\}
\vskip 1.0em
\vskip 3.0em
\end{center}
\begin{abstract}
We study the wave equations with various spins on the background of the general spherically symmetric spacetime.
We obtain the unified expression of the Teukolsky-like master equations and the corresponding radial equations with the general spins.
We also discuss the gauge dependence in the gravitational-wave equations, which have appeared in the previous studies.

\end{abstract}
\end{titlepage}

\section{Introduction}
The study on the gravitational waves becomes much more important subject in cosmology than before, since the
direct observation of gravitational waves by LIGO and VIRGO~\cite{Abbott:2016blz}.
The gravitational waves can be desribed as a solution to the Einstein equation, where the metric is perturbed
around a certain background.
When one chooses the background as the black hole spacetime, one can for example consider the gravitational-wave radiation
from a relatively light object rotating around the black hole, which can be studied by
the black-hole perturbation theory \cite{Mino:1997bx}.
On the other hand, when the mass of the object is not so light, the contribution to the background spacetime from this object
may not be negligible. Then one has to modify the background, or has to use some other methods,
such as the post-Newtonian approximation \cite{PoissonWill}. One systematic method to modify the background is proposed as
the effective one-body (EOB) dynamics \cite{Buonanno:2000ef,Damour:2016gwp}.

The background of the EOB dynamics is deformed from the black hole spacetime, and then may not satisfy the vacuum Einstein equation.
Motivated by this, we will consider the general spherically symmetric spacetime as a simple example of the background, which is not
necessarily the vacuum.
A particular form of the background indeed appears in the EOB dynamics for the spinless binary system \cite{Buonanno:2000ef,Damour:2016gwp}.
The spherically symmetric spacetime satisfies the Petrov type D condition,
which has played the key role to derive the gravitational-wave equation in the previous studies \cite{Jing:2021ahx, Jing:2022vks, Guo:2023niy}
using the Newman-Penrose formalism \cite{Newman:1961qr}, as for the Teukolsky equation \cite{Teukolsky:1973ha} in the vacuum case.
The advantage to use the Newman-Penrose formalism is that the role of the Einstein equation is rather restrictive, which is just used
to relate the Ricci tensor to the energy-momentum tensor. Then the extension to the non-vacuum case of relatively easier.
It has been found that in order to obtain the decoupled wave equation, one has to take the
gauge condition such that some of the coupled degrees of freedom are taken to vanish \cite{Jing:2021ahx, Jing:2022vks, Guo:2023niy}.
So far two types of the gravitational-wave equation are proposed \cite{Jing:2022vks, Guo:2023niy} due to the difference of the gauge conditions.

In this paper we will study the massless wave equations with different spins (helicities, more precisely), in other words,
the (massless) Klein-Gordon equation, the Weyl equation and the Maxwell equation on the same background spacetime.
In order to avoid the complexities, we will first give the unified expression of those equations in the level of the equation
which consists of the Newman-Penrose quantities \textbf{as in \cite{Harris:2003eg, Vagenas:2020bys, Arbey:2021jif, Li:2011za}},
and then obtain the explicit wave equation and the ordinary differential equation for the radial coordinate.

The remaining of this paper is organized as follows: In section 2, we will introduce our parametrization of the
background of the spherically symmetric spacetime, and show the quantities in the Newman-Penrose formalism.
In section 3, we will see the wave equation for spin $0$, $\pm 1/2$, $\pm 1$ and $\pm 2$ on this background.
Then we will give the unified expression for those equations with the general spin $s$.
We will also obtain the explicit Teukolsky-like master equation and the corresponding radial equation.
In section 4, we will discuss the gauge dependence in the gravitational-wave equations proposed in the previous studies.
Section 5 is devoted to the summary and the discussion.

\section{Background metric and tetrads}

We will consider the general spherically symmetric background as
\begin{gather}
ds^{2}=\mathcal{A}(r)dt^{2}-\mathcal{B}(r)dr^{2}-\mathcal{C}(r)r^{2}(d\theta^{2}+\sin^{2}\theta d\varphi^{2}).
\label{metric1}
\end{gather}
By choosing the radial coordinate $r$ appropriately, we can set $\mathcal{C}(r)=1$, which corresponds to the standard coordinate.
We define $\mathcal{D}(r)=\sqrt{\mathcal{A}(r)\mathcal{B}(r)}$, and then the background metric \eqref{metric1} becomes
\begin{gather}
ds^{2}=\mathcal{A}(r)dt^{2}-\frac{\mathcal{D}(r)^{2}}{\mathcal{A}(r)}dr^{2}-r^{2}(d\theta^{2}+\sin^{2}\theta d\varphi^{2}).
\label{metric2}
\end{gather}
The null tetrads corresponding to the metric \eqref{metric2} are taken as
\begin{align}
l=l_{\mu}^{A}dx^{\mu}&=dt-\frac{\mathcal{D}(r)}{\mathcal{A}(r)}dr, &
n=n_{\mu}^{A}dx^{\mu}&=\frac{\mathcal{A}(r)}{2}dt+\frac{\mathcal{D}(r)}{2}dr, \notag\\
m=m_{\mu}^{A}dx^{\mu}&=-\frac{r}{\sqrt{2}}(d\theta+i\sin\theta d\varphi), &
\bar{m}=\bar{m}_{\mu}^{A}dx^{\mu}&=-\frac{r}{\sqrt{2}}(d\theta-i\sin\theta d\varphi),
\label{tetrads1}
\end{align}
which satisfy
\begin{gather}
ds^{2}=2ln-2m\bar{m}.
\label{decomp}
\end{gather}
Here the superscript (or the subscript) $A$ is used as the symbol of the background quantities \cite{Teukolsky:1973ha}.
On the other hand, for the perturbation quantities of the gravitational field, we put the superscipt $B$, which will appear later.

From the tetrad basis \eqref{tetrads1}, one can compute the spin coefficients, the components
of the Ricci tensor and the Weyl scalars as
\begin{gather}
\kappa^{A}=\nu^{A}=\sigma^{A}=\lambda^{A}=\pi^{A}=\tau^{A}=\epsilon^{A}=0, \label{bg11}\\
\rho^{A}=-\frac{1}{r\mathcal{D}}, \quad \mu^{A}=-\frac{\mathcal{A}}{2r\mathcal{D}},
\quad \gamma^{A}=\frac{\mathcal{A}'}{4\mathcal{D}}, \quad \alpha^{A}=-\beta^{A}=-\frac{\cot\theta}{2\sqrt{2}r}, \label{bg12}\\
\Phi_{01}^{A}=\Phi_{10}^{A}=\Phi_{02}^{A}=\Phi_{20}^{A}=\Phi_{12}^{A}=\Phi_{21}^{A}=0, \\
\Phi_{00}^{A}=\frac{\mathcal{D}'}{r\mathcal{D}^{3}}, \quad
\Phi_{22}^{A}=\frac{\mathcal{A}^{2}\mathcal{D}'}{4rD^{3}}, \\
\Phi_{11}^{A}=\frac{1}{8r^{2}\mathcal{D}^{3}}\Bigl[2\mathcal{D}^{3}-2\mathcal{A}\mathcal{D}
-r^{2}(\mathcal{A}'\mathcal{D}'-\mathcal{A}''\mathcal{D})\Bigr], \\
\Lambda^{A}=-\frac{1}{24r^{2}\mathcal{D}^{3}}\Bigl[-2\mathcal{D}^{3}-r^{2}\mathcal{A}'\mathcal{D}'
+2\mathcal{A}(\mathcal{D}-2r\mathcal{D}')+r\mathcal{D}(4\mathcal{A}'+r\mathcal{A}'')\Bigr], \\
\Psi_{0}^{A}=\Psi_{1}^{A}=\Psi_{3}^{A}=\Psi_{4}^{A}=0, \label{typeD}\\
\Psi_{2}^{A}=\frac{1}{12r^{2}\mathcal{D}^{3}}\Bigl[2(\mathcal{A}\mathcal{D}+r\mathcal{A}\mathcal{D}'-\mathcal{D}^{3})-r\mathcal{A}'(2\mathcal{D}+r\mathcal{D}')+r^{2}\mathcal{A}''\mathcal{D}\Bigr], \label{bg18}
\end{gather}
where we follow the notations of Newman-Penrose \cite{Newman:1961qr} and Pirani \cite{Pirani}.
The same notation is also used in Teukolsky \cite{Teukolsky:1973ha}.
The prime denotes the derivative with respect to $r$.
The equation \eqref{typeD} implies that the background belongs the Petrov type D, which will be important to derive the wave equation
on this background. As an special case, for $\mathcal{D}(r)=1$ we have
\begin{gather}
\Phi_{00}=\Phi_{22}=0,
\label{0022}
\end{gather}
and the nonvanishing quantities in the background are simplified as
\begin{gather}
\rho^{A}=-\frac{1}{r}, \quad \mu^{A}=-\frac{\mathcal{A}}{2r}, \quad \gamma^{A}=\frac{\mathcal{A}'}{4},
\quad \alpha^{A}=-\beta^{A}=-\frac{\cot\theta}{2\sqrt{2}r}, \label{bg21}\\
\Phi_{11}^{A}=\frac{1}{8r^{2}}(2-2\mathcal{A}+r^{2}\mathcal{A}''), \\
\Lambda^{A}=-\frac{1}{24r^{2}}(-2+2\mathcal{A}+4r\mathcal{A}'+r^{2}\mathcal{A}''), \\
\Psi_{2}^{A}=\frac{1}{12r^{2}}(-2+2\mathcal{A}-2r\mathcal{A}'+r^{2}\mathcal{A}''), \label{bg24}
\end{gather}
We also note that when we choose $\mathcal{A}=1-2M/r$ and $\mathcal{D}=1$, the background reduces to
the Schwarzschild spacetime, where $M$ is the mass of the Schwarzschild black hole.

\section{Wave equations with various spins}

Here we will consider the wave equation with the various spin $s$ on the background specified by \eqref{bg11}--\eqref{bg18} in the previous section.
Namely, the Klein-Gordon equation ($s=0$), the Weyl equation ($s=\pm 1/2$), the Maxwell equation ($s=\pm 1$)
and the equation from the Newman-Penrose formalism ($s=\pm 2$) under the probe (test field) approximation,
which means the back reactions from the matters and the electromagnetic fields to the gravitational background are assumed to be negligible.
For simplicity, we will also assume that the fields are massless and are minimally coupled to the background of the gravitational field,
unless it is specified.

\subsection{spin \text{$0$}}

The massless Klein-Gordon equation in the gravitational background is
\begin{gather}
\Box\phi=\nabla_{\mu}(g^{\mu\nu}\partial_{\nu}\phi)=\frac{1}{\sqrt{-g}}\partial_{\mu}(\sqrt{-g}g^{\mu\nu}\partial_{\nu}\phi)=T,
\label{KG1}
\end{gather}
where $T$ is the source.
$\nabla_{\mu}$ is the covariant derivative with respect to the curved spacetime (not for the local Lorentz transformation),
of which the projection by the null tetrads gives
\begin{gather}
D^{A}=l_{A}^{\mu}\nabla_{\mu}, \quad \Delta^{A}=n_{A}^{\mu}\nabla_{\mu}, \quad \delta^{A}=m_{A}^{\mu}\nabla_{\mu}, \quad
\bar{\delta}^{A}=\bar{m}_{A}^{\mu}\nabla_{\mu}.
\label{diffop}
\end{gather}
By decomposing $g^{\mu\nu}$ in terms of the null tetrads, \eqref{KG1} can be rewritten as
\begin{gather}
\left[(\Delta-2\gamma+2\mu)D+(D-2\rho)\Delta-(\bar{\delta}-2\alpha)\delta-(\delta-2\alpha)\bar{\delta}\right]^{A}\phi=T,
\label{KG2}
\end{gather}
where the superscript $A$ outside of the parentheses denotes that all the quantities and the operators inside the parentheses
are the background ones, and we have used the relation for the spin coefficitents as
\begin{gather}
\nabla_{\mu}l_{A}^{\mu}=-2\rho^{A}, \quad \nabla_{\mu}n_{A}^{\mu}=-2\gamma^{A}+2\mu^{A}, \quad
\nabla_{\mu}m_{A}^{\mu}=\nabla_{\mu}\bar{m}_{A}^{\mu}=-2\alpha^{A}.
\end{gather}
For later convenience, we will rewrite \eqref{KG2} such that the order of the differential oprerators is rearranged,
which satisfy the commutation relations
\begin{align}
\Delta^{A} D^{A}-D^{A} \Delta^{A}=2\gamma^{A} D^{A}, \quad
\bar{\delta}^{A}\delta^{A}-\delta^{A}\bar{\delta}^{A}=2\alpha^{A}(\delta-\bar{\delta})^{A}.
\label{com0}
\end{align}
We will also use
\begin{gather}
\Delta^{A}\rho^{A}=(2\gamma-\mu)^{A}\rho^{A}-\Psi_{2}^{A}-2\Lambda^{A},
\label{NP1}
\end{gather}
which is from the background part of the one of the Newman-Penrose equation.
Then \eqref{KG2} can be rewritten as
\begin{gather}
\left[(\Delta-2\gamma+\mu)(D-\rho)-(\bar{\delta}-2\alpha)\delta-\Psi_{2}-2\Lambda\right]^{A}\phi=\frac{1}{2}T,
\label{KG3}
\end{gather}
We note that the last term $-2\Lambda^{A}\phi$ in the right hand side is responsible to the minimal coupling.
If we consider the curvature coupling, there is the contribution $cR^{A}\phi=24c\Lambda^{A}\phi$, where $c$ is constant.

\subsection{spin \text{$\pm 1/2$}}

The massless Dirac equation can be decomposed into two Weyl equations.
For the positive chirality part, the Weyl equation in the gravitational background is
\begin{align}
(\bar{\delta}-\alpha)^{A}\chi_{0}-(D-\rho)^{A}\chi_{1}=0,
\label{weyl1}
\\
(\Delta-\gamma+\mu)^{A}\chi_{0}-(\delta-\alpha)^{A}\chi_{1}=0,
\label{weyl2}
\end{align}
where $\chi_{0}$ and $\chi_{1}$ are the components of the Weyl spinor.
One can eliminate $\chi_{0}$ using the following commutation relation:
\begin{align}
&\left[\Delta+p\gamma-(q-1)\mu\right]^{A}(\bar{\delta}+p\alpha)^{A}
-\left(\bar{\delta}+p\alpha\right)^{A}(\Delta+p\gamma-q\mu)^{A} \notag\\
&=\nu^{A}D^{A}-\lambda^{A}\delta^{A}+p\left(\alpha\lambda+\rho\nu-\Psi_{3}\right)^{A}
+q\left(-D\nu+\delta\lambda-4\alpha\lambda+2\Psi_{3}\right)^{A} \notag\\
&=0, \label{com1}
\end{align}
where $p$ and $q$ are arbitrary constants and we have just used $\nu^{A}=\lambda^{A}=\Psi_{3}^{A}=0$ for the last equality.
Hence \eqref{com1} holds not only on the vacuum, but also on the background specified by \eqref{bg11}--\eqref{bg18}.
We will obtain the wave equation for $\chi_{1}$ in a similar way with the method
used to derive the Teukolsky equation \cite{Teukolsky:1973ha}.
We operate $(\Delta-\gamma+2\mu)^{A}$ on \eqref{weyl1} and $(\bar{\delta}-\alpha)^{A}$ on \eqref{weyl2}, respectively.
We then make the difference of them. The terms with $\chi_{0}$ are canceled from \eqref{com1} with $p=q=-1$,
and the remaining part is
\begin{gather}
\bigl[(\Delta-\gamma+2\mu)(D-\rho)-(\bar{\delta}-\alpha)(\delta-\alpha)\bigr]^{A}\chi_{1}=0,
\label{chi1}
\end{gather}
which gives the wave equation for $s=-1/2$. In a similar way, the wave equation of $\chi_{0}$ (for $s=1/2$) is obtained as
\begin{gather}
\bigl[(D-2\rho)(\Delta-\gamma+\mu)-(\delta-\alpha)(\bar{\delta}-\alpha)\bigr]\chi_{0}=0.
\label{chi0-1}
\end{gather}
We note that \eqref{chi1} and \eqref{chi0-1} take the same form as the vacuum case \cite{Teukolsky:1973ha}.
For later convenience, we rewrite \eqref{chi0-1} as in the previous subsection. We will use \eqref{com0}, \eqref{NP1} and
\begin{align}
D^{A}\gamma^{A}&=\Psi_{2}^{A}+\Phi_{11}^{A}-\Lambda^{A},
\label{NP2}
\\
(\delta+\bar{\delta})^{A}\alpha^{A}&=\mu^{A}\rho^{A}+4(\alpha^{A})^{2}-\Psi_{2}^{A}+\Phi_{11}^{A}+\Lambda^{A},
\label{NP3}
\\
D^{A}\mu^{A}&=\mu^{A}\rho^{A}+\Psi_{2}^{A}+2\Lambda^{A}.
\label{NP4}
\end{align}
Then \eqref{chi0-1} can be rewritten as
\begin{gather}
\bigl[(\Delta-3\gamma+\mu)(D-2\rho)-(\bar{\delta}-3\alpha)(\delta+\alpha)-3\Psi_{2}\bigr]^{A}\chi_{0}=0,
\label{chi0-2}
\end{gather}

\subsection{spin \text{$\pm 1$}}

The Maxwell equation in the gravitational background is
\begin{align}
(D-2\rho)^{A}\phi_{1}-(\bar{\delta}-2\alpha)^{A}\phi_{0}&=J_{l},
\label{max1}
\\
\delta^{A}\phi_{1}-(\Delta+\mu-2\gamma)^{A}\phi_{0}&=J_{m},
\label{max2}
\\
(D-\rho)^{A}\phi_{2}-\bar{\delta}^{A}\phi_{1}&=J_{\bar{m}},
\label{max3}
\\
(\delta-2\alpha)^{A}\phi_{2}-(\Delta+2\mu)\phi_{1}&=J_{n},
\label{max4}
\end{align}
where $\phi_{0}$, $\phi_{1}$ and $\phi_{2}$ are complex, and are constructed from the field strength (the Faraday tensor) $F_{\mu\nu}$ as%
\footnote{As in the vacuum case, we will consider the wave equation just for $\phi_{0}$ and $\phi_{2}$, from which $\phi_{1}$ can be obtained using
\eqref{max1}--\eqref{max4}. }
\begin{gather}
\phi_{0}=F_{\mu\nu}l^{\mu}m^{\nu}, \quad
\phi_{1}=\frac{1}{2}F_{\mu\nu}(l^{\mu}n^{\nu}+\bar{m}^{\mu}m^{\nu}), \quad
\phi_{2}=F_{\mu\nu}\bar{m}^{\mu}n^{\nu}.
\end{gather}
$J_{l}$, $J_{n}$, $J_{m}$ and $J_{\bar{m}}$ are the projection of the current $J^{\mu}$ by the null tetrads as $J_{l}=J^{\mu}l_{\mu}^{A}$, etc.
From \eqref{max3} and \eqref{max4}, one can make the wave equation for $\phi_{2}$ by a similar procedure used in the previous subsection.
Using the commutation relation \eqref{com1} with $p=0$ and $q=-2$, we have
\begin{gather}
\bigl[(\Delta+3\mu)(D-\rho)-\bar{\delta}(\delta-2\alpha)\bigr]^{A}\phi_{2}=J_{2},
\label{phi2}
\end{gather}
where $J_{2}$ is defined by
\begin{gather}
J_{2}=(\Delta+3\mu)^{A}J_{\bar{m}}-\bar{\delta}J_{n}.
\end{gather}
In a similar way, one can obtain the wave equation for $\phi_{0}$ from \eqref{max1} and \eqref{max2} as
\begin{gather}
\bigl[(D-3\rho)(\Delta-2\gamma+\mu)-\delta(\bar{\delta}-2\alpha)\bigr]^{A}\phi_{0}=J_{0},
\label{phi0-1}
\end{gather}
where $J_{0}$ is defined by
\begin{gather}
J_{0}=\delta J_{l}-(D-3\rho)J_{m}.
\end{gather}
We note that \eqref{phi2} and \eqref{phi0-1} take the same form as the vacuum case \cite{Teukolsky:1973ha}
For later convenience, we rewrite \eqref{phi0-1} using \eqref{com0}, \eqref{NP1} and \eqref{NP2}--\eqref{NP4} as
\begin{gather}
\bigl[(\Delta-4\gamma+\mu)(D-3\rho)-(\bar{\delta}-4\alpha)(\delta+2\alpha)-6\Psi_{2}\bigr]^{A}\phi_{0}=J_{0},
\label{phi0-2}
\end{gather}

\subsection{spin \text{$\pm 2$}}

The wave equation for the spin $\pm 2$ is obtained from the perturbed Einstein equation or the perturbed Newman-Penrose equation.
In the previous paper \cite{Guo:2023niy} we have studied and have obtained the wave equations,
and in the next section we will also revisit the derivation in order to discuss the gauge dependence.
Then we here just give the result of the equations.

We will here take the gauge such that
the following quantities vanish \cite{Guo:2023niy}:
\begin{gather}
\lambda^{B}=\sigma^{B}=0,
\label{gauge01}
\\
(\bar{\delta}-\bar{\tau}+2\alpha+2\bar{\beta})^{B}\Phi_{22}^{A}=0,
\label{gauge02}
\\
(\delta+\bar{\pi}-2\bar{\alpha}-2\beta)^{B}\Phi_{00}^{A}=0.
\label{gauge03}
\end{gather}
Under the above gauge, the wave equation for the perturbation part of the Weyl scalar $\Psi_{4}^{B}$ is
\begin{gather}
\bigl[(\Delta+2\gamma+5\mu)(D-\rho)-(\bar{\delta}+2\alpha)(\delta-4\alpha)-3\Psi_{2}+2\Phi_{11}\bigr]^{A}\Psi_{4}^{B}=T_{4},
\label{GW04}
\end{gather}
where the source $T_{4}$ is defined by
\begin{align}
T_{4}&=(\Delta+2\gamma+5\mu)^{A}\bigl[(\bar{\delta}+2\alpha)^{A}\Phi_{21}^{B}-(\Delta+\mu)^{A}\Phi_{20}^{B}\bigr]
\notag\\
&\qquad {}-(\bar{\delta}+2\alpha)^{A}\bigl[\bar{\delta}\Phi_{22}^{B}-(\Delta+2\gamma+2\mu)^{A}\Phi_{21}^{B})\bigr].
\end{align}
In a similar way, one can also obtain the wave equation for $\Psi_{0}^{B}$ as
\begin{gather}
\bigl[(D-5\rho)(\Delta-4\gamma+\mu)-(\delta+2\alpha)(\bar{\delta}-4\alpha)-3\Psi_{2}+2\Phi_{11}\bigr]^{A}\Psi_{0}^{B}=T_{0},
\label{GW00-1}
\end{gather}
where the source $T_{0}$ is defined by
\begin{align}
T_{0}&=(\delta+2\alpha)^{A}\bigl[(D-2\rho)\Phi_{01}^{B}-\delta^{A}\Phi_{00}^{B}\bigr]
\notag\\
&\qquad {}-(D-5\rho)^{A}\bigl[(D-\rho)^{A}\Phi_{02}^{B}-(\delta+2\alpha)^{A}\Phi_{01}^{B}\bigr].
\end{align}
For later convenience, we rewrite \eqref{GW00-1} using \eqref{com0}, \eqref{NP1} and \eqref{NP2}--\eqref{NP4} as
\begin{gather}
\bigl[(\Delta-6\gamma+\mu)(D-5\rho)-(\bar{\delta}-6\alpha)(\delta+4\alpha)-15\Psi_{2}+2\Phi_{11}\bigr]^{A}\Psi_{0}^{B}=T_{0}.
\label{GW00-2}
\end{gather}

\subsection{unified wave equation for general spin \text{$s$}}

We can unify the above wave equations \eqref{KG3}, \eqref{chi1}, \eqref{chi0-2}, \eqref{phi2}, \eqref{phi0-2}, \eqref{GW04} and \eqref{GW00-2},
and can express the one for the general spin $s$ as
\begin{align}
&\biggl\{\bigl[\Delta-2(1+s)\gamma+(1-s+|s|)\mu\bigr]\bigl[D-(1+s+|s|)\rho\bigr]-\bigl[\bar{\delta}-2(1+s)\alpha\bigr](\delta+2s\alpha)
\notag\\
&\qquad {}-(1+3s+2s^{2})\Psi_{2}+\frac{1}{3}(|s|-3|s|^{2}+2|s|^{3})\Phi_{11}-2\delta_{s}\Lambda\biggr\}^{A}\tilde{\psi}_{(s)}=\tilde{T}_{(s)},
\label{unified1}
\end{align}
where we have collectively denoted the fields and the sources as
\begin{gather}
\tilde{\psi}_{(s)}=
\begin{cases}
\Psi_{4}^{B} & \text{for $s=-2$}
\\
\phi_{2} & \text{for $s=-1$}
\\
\chi_{1} & \text{for $s=-1/2$}
\\
\phi & \text{for $s=0$}
\\
\chi_{0} & \text{for $s=1/2$}
\\
\phi_{0} & \text{for $s=1$}
\\
\Psi_{0}^{B} & \text{for $s=2$}
\end{cases},
\qquad
\tilde{T}_{(s)}=
\begin{cases}
T_{4} & \text{for $s=-2$}
\\
J_{2} & \text{for $s=-1$}
\\
0 & \text{for $s=-1/2$}
\\
T/2 & \text{for $s=0$}
\\
0 & \text{for $s=1/2$}
\\
J_{0} & \text{for $s=1$}
\\
T_{0} & \text{for $s=2$}
\end{cases}.
\end{gather}
$\delta_{s}$ is defined by
\begin{gather}
\delta_{s}=
\begin{cases}
1\quad\text{for $s=0$}
\\
0\quad\text{otherwise}.
\end{cases}
\end{gather}
We will rewrite \eqref{unified1} to the slightly simpler form by the following redefinition;
\begin{gather}
\psi_{(s)}=\exp\bigl[(|s|-s)f\bigr]\tilde{\psi}_{(s)}, \quad T_{(s)}=\exp\bigl[(|s|-s)f\bigr]\tilde{T}_{(s)},
\label{redef}
\end{gather}
where $f$ is a function of $r$, which will be determined soon. By substituting \eqref{redef} into \eqref{unified1}, we have
\begin{align}
&\biggl\{\bigl[\Delta-2(1+s)\gamma+\mu+(|s|-s)(\mu-\Delta f)\bigr]\bigl[D-(1+2s)\rho-(|s|-s)(\rho+Df)\bigr]
\notag\\
&\qquad {}-\bigl[\bar{\delta}-2(1+s)\alpha\bigr](\delta+2s\alpha)-(1+3s+2s^{2})\Psi_{2}
\notag\\
&\qquad {}+\frac{1}{3}(|s|-3|s|^{2}+2|s|^{3})\Phi_{11}-2\delta_{s}\Lambda\biggr\}^{A}\psi_{(s)}=T_{(s)}.
\label{unified1-1}
\end{align}
One can find that $\mu^{A}-\Delta^{A} f$ and $\rho^{A}+D^{A}f$ can simultaneously vanish by choosing $f=\ln r$, namely%
\footnote{The additive integration constant for $f$ is irrelevant here, and is chosen in convenience. }
\begin{gather}
\psi_{(s)}= r^{|s|-s}\tilde{\psi}_{(s)},\quad T_{(s)}= r^{|s|-s}\tilde{T}_{(s)}.
\label{redef1}
\end{gather}
Then \eqref{unified1-1} is simplified as
\begin{align}
&\biggl\{\bigl[\Delta-2(1+s)\gamma+\mu\bigr]\bigl[D-(1+2s)\rho\bigr]-\bigl[\bar{\delta}-2(1+s)\alpha\bigr](\delta+2s\alpha)
\notag\\
&\qquad {}-(1+3s+2s^{2})\Psi_{2}+\frac{1}{3}(|s|-3|s|^{2}+2|s|^{3})\Phi_{11}-2\delta_{s}\Lambda\biggr\}^{A}\psi_{(s)}=T_{(s)}.
\label{unified2}
\end{align}
The advantage of the above form is that in the vacuum $\Phi_{11}^{A}=\Lambda^{A}=0$,
the equation \eqref{unified2} just depends on $s$, but not $|s|$.
The same transformation \eqref{redef1} has been performed in the vacuum case as well \cite{Teukolsky:1973ha}.
\textbf{Similar equations as \eqref{unified1} and \eqref{unified2} have been studied in
\cite{Harris:2003eg, Vagenas:2020bys, Arbey:2021jif, Li:2011za}, which are however
considered for positive and negative $s$ separately, or restricted to positive $s$. Here we have obtained the completely
unified expression for both positive and negative $s$. Moreover, we have found the contribution of $\Phi_{11}^{A}$ and $\Lambda^{A}$
to the wave equation. }

By substituting the background, the explicit form of the unified wave equation \eqref{unified2} is
\begin{align}
&\frac{r^{2}}{\mathcal{A}}\frac{\partial^{2}\psi_{(s)}}{\partial t^{2}}
-\frac{1}{\mathcal{D}(r^{2}\mathcal{A})^{s}}\frac{\partial}{\partial r}
\left[\frac{(r^{2}\mathcal{A})^{s+1}}{\mathcal{D}}\frac{\partial\psi_{(s)}}{\partial r}\right]
-\frac{1}{\sin\theta}\frac{\partial}{\partial\theta}\left(\sin\theta\frac{\partial\psi_{(s)}}{\partial\theta}\right)
-\frac{1}{\sin^{2}\theta}\frac{\partial^{2}\psi_{(s)}}{\partial\varphi^{2}}
\notag\\
&\qquad {}+\left(\frac{2sr}{\mathcal{D}}-\frac{sr^{2}\mathcal{A}'}{\mathcal{A}\mathcal{D}}\right)\frac{\partial\psi_{(s)}}{\partial t}
-\frac{2is\cot\theta}{\sin\theta}\frac{\partial\psi_{(s)}}{\partial\varphi}
+\biggl[s^{2}\cot^{2}\theta-s-\frac{s(2s+1)r\mathcal{A}\mathcal{D}'}{\mathcal{D}^{3}}
\notag\\
&\qquad {}+\frac{1}{3}(1-\delta_{s}+3s+2s^{2})\left(1-\frac{\mathcal{A}}{\mathcal{D}^{2}}-\frac{2r\mathcal{A}'}{\mathcal{D}^{2}}
+\frac{2r\mathcal{A}\mathcal{D}'}{\mathcal{D}^{3}}+\frac{r^{2}\mathcal{A'}\mathcal{D}'}{2\mathcal{D}^{3}}
-\frac{r^{2}\mathcal{A}''}{2\mathcal{D}^{2}}
\right)
\notag\\
&\qquad {}+\frac{1}{6}(|s|-3|s|^{2}+2|s|^{3})\left(
1-\frac{\mathcal{A}}{\mathcal{D}^{2}}-\frac{r^{2}\mathcal{A}'\mathcal{D}'}{2\mathcal{D}^{3}}+\frac{r^{2}\mathcal{A}''}{2\mathcal{D}^{2}}
\right)
\biggr]\psi_{(s)}=2r^{2}T_{(s)}.
\label{PDE1}
\end{align}
We note that \eqref{PDE1} with $s=\pm 2$ looks different from the equation obtained in our previous study \cite{Guo:2023niy}.
However this is just due to the difference of the transformation \eqref{redef1}, and they are equivalent.
We also note that in the case of $\mathcal{A}=1-2M/r$ and $\mathcal{D}=1$, \eqref{PDE1} reduces to the Teukolsky master equation
with the spin $s$ on the background of the Schwarzschild spacetime.
In the case of $\mathcal{D}=1$, The above is simplified as
\begin{align}
&\frac{r^{2}}{\mathcal{A}}\frac{\partial^{2}\psi_{(s)}}{\partial t^{2}}
-\frac{1}{(r^{2}\mathcal{A})^{s}}\frac{\partial}{\partial r}
\left[(r^{2}\mathcal{A})^{s+1}\frac{\partial\psi_{(s)}}{\partial r}\right]
-\frac{1}{\sin\theta}\frac{\partial}{\partial\theta}\left(\sin\theta\frac{\partial\psi_{(s)}}{\partial\theta}\right)
\notag\\
&\qquad {}-\frac{1}{\sin^{2}\theta}\frac{\partial^{2}\psi_{(s)}}{\partial\varphi^{2}}
+\left(2sr-\frac{sr^{2}\mathcal{A}'}{\mathcal{A}}\right)\frac{\partial\psi_{(s)}}{\partial t}
-\frac{2is\cot\theta}{\sin\theta}\frac{\partial\psi_{(s)}}{\partial\varphi}
\notag\\
&\qquad {}+\biggl[s^{2}\cot^{2}\theta-s
+\frac{1}{3}(1-\delta_{s}+3s+2s^{2})\left(1-\mathcal{A}-2r\mathcal{A}'-\frac{1}{2}r^{2}\mathcal{A}''
\right)
\notag\\
&\qquad\qquad {}+\frac{1}{6}(|s|-3|s|^{2}+2|s|^{3})\left(1-\mathcal{A}+\frac{1}{2}r^{2}\mathcal{A}''
\right)
\biggr]\psi_{(s)}=2r^{2}T_{(s)}.
\label{PDE2}
\end{align}
First we consider the homogeneous case. the equation allows the separation of the variables and we assume the product form of the solution as
\begin{gather}
\psi_{(s)}=e^{-i\omega t}e^{im\varphi}R(r)S(\theta),
\end{gather}
where $\omega$ is the frequency of the waves and $m$ is constant.
Then the separated equations are
\begin{align}
&\frac{1}{\mathcal{D}(r^{2}\mathcal{A})^{s}}\frac{d}{dr}\left[\frac{(r^{2}\mathcal{A})^{s+1}}{\mathcal{D}}\frac{dR}{dr}\right]
+\biggl[\frac{r^{2}\omega^{2}}{\mathcal{A}}+i\omega\left(\frac{2sr}{\mathcal{D}}-\frac{sr^{2}\mathcal{A}'}{\mathcal{A}\mathcal{D}}\right)
+\frac{s(2s+1)r\mathcal{A}\mathcal{D}'}{\mathcal{D}^{3}}
\notag\\
&\qquad {}-\frac{1}{3}(1-\delta_{s}+3s+2s^{2})\left(1-\frac{\mathcal{A}}{\mathcal{D}^{2}}-\frac{2r\mathcal{A}'}{\mathcal{D}^{2}}
+\frac{2r\mathcal{A}\mathcal{D}'}{\mathcal{D}^{3}}+\frac{r^{2}\mathcal{A'}\mathcal{D}'}{2\mathcal{D}^{3}}
-\frac{r^{2}\mathcal{A}''}{2\mathcal{D}^{2}}
\right)
\notag\\
&\qquad {}-\frac{1}{6}(|s|-3|s|^{2}+2|s|^{3})\left(
1-\frac{\mathcal{A}}{\mathcal{D}^{2}}-\frac{r^{2}\mathcal{A}'\mathcal{D}'}{2\mathcal{D}^{3}}+\frac{r^{2}\mathcal{A}''}{2\mathcal{D}^{2}}
\right)
-\boldsymbol{\lambda}_{(s)}\biggr]R=0,
\label{radial}
\\
&\frac{1}{\sin\theta}\frac{d}{d\theta}\left(\sin\theta\frac{dS}{d\theta}\right)
+\left(
-\frac{m^{2}}{\sin^{2}\theta}-\frac{2sm\cot\theta}{\sin\theta}-s^{2}\cot^{2}+s+\boldsymbol{\lambda}_{(s)}
\right)S=0,
\label{angular}
\end{align}
where $\boldsymbol{\lambda}_{(s)}$ is the separation constant. From \eqref{angular}, one can find that $S(\theta)e^{im\varphi}$
coincides with the spin-weighted spherical harmonics ${}_{s}Y_{lm}(\theta,\varphi)$ with the spin $s$,
where $l$ and $m$ take the values of
\begin{gather}
l=|s|,\ |s|+1,\ |s|+2,\ \ldots, \quad m= -l,\ -l+1,\ \ldots,\ l-1,\ l,
\end{gather}
respectively.
$\boldsymbol{\lambda}_{(s)}$ becomes the eigenvalue of ${}_{s}Y_{lm}(\theta,\varphi)$, which is given by
\begin{gather}
\boldsymbol{\lambda}_{(s)}=(l-s)(l+s+1).
\end{gather}
For the nonhomogeneous case, we expand $\psi_{(s)}$ and $T_{(s)}$ in terms of ${}_{s}Y_{lm}(\theta,\varphi)$ as
\begin{align}
\psi_{(s)}&=\int d\omega \sum_{l,m}R^{(s)}_{l\omega}(r){}_{s}Y_{lm}(\theta,\varphi)e^{-i\omega t},
\\
-2r^{2}T_{(s)}&=\int d\omega \sum_{l,m}G^{(s)}_{l\omega}(r){}_{s}Y_{lm}(\theta,\varphi)e^{-i\omega t}.
\end{align}
Then $R^{(s)}_{l\omega}(r)$ satisfies
\begin{align}
&\frac{1}{\mathcal{D}(r^{2}\mathcal{A})^{s}}\frac{d}{dr}\left[\frac{(r^{2}\mathcal{A})^{s+1}}{\mathcal{D}}
\frac{dR^{(s)}_{l\omega}}{dr}\right]
+\biggl[\frac{r^{2}\omega^{2}}{\mathcal{A}}+i\omega\left(\frac{2sr}{\mathcal{D}}-\frac{sr^{2}\mathcal{A}'}{\mathcal{A}\mathcal{D}}\right)
+\frac{s(2s+1)r\mathcal{A}\mathcal{D}'}{\mathcal{D}^{3}}
\notag\\
&\qquad {}-\frac{1}{3}(1-\delta_{s}+3s+2s^{2})\left(1-\frac{\mathcal{A}}{\mathcal{D}^{2}}-\frac{2r\mathcal{A}'}{\mathcal{D}^{2}}
+\frac{2r\mathcal{A}\mathcal{D}'}{\mathcal{D}^{3}}+\frac{r^{2}\mathcal{A'}\mathcal{D}'}{2\mathcal{D}^{3}}
-\frac{r^{2}\mathcal{A}''}{2\mathcal{D}^{2}}
\right)
\notag\\
&\qquad {}-\frac{1}{6}(|s|-3|s|^{2}+2|s|^{3})\left(
1-\frac{\mathcal{A}}{\mathcal{D}^{2}}-\frac{r^{2}\mathcal{A}'\mathcal{D}'}{2\mathcal{D}^{3}}+\frac{r^{2}\mathcal{A}''}{2\mathcal{D}^{2}}
\right)
-\boldsymbol{\lambda}_{(s)}\biggr]R^{(s)}_{l\omega}=G^{(s)}_{l\omega}.
\label{radial2}
\end{align}
We again note that in the case of $\mathcal{A}=1-2M/r$ and $\mathcal{D}=1$, \eqref{radial2} reduces to the Teukolsky radial equation
with the spin $s$ on the background of the Schwarzschild spacetime.
In the case of $\mathcal{D}=1$, \eqref{radial2} reduces to
 \begin{align}
&\frac{1}{(r^{2}\mathcal{A})^{s}}\frac{d}{dr}\left[(r^{2}\mathcal{A})^{s+1}
\frac{dR^{(s)}_{l\omega}}{dr}\right]
+\biggl[\frac{r^{2}\omega^{2}}{\mathcal{A}}+i\omega\left(2sr-\frac{sr^{2}\mathcal{A}'}{\mathcal{A}}\right)
\notag\\
&\qquad {}-\frac{1}{3}(1-\delta_{s}+3s+2s^{2})\left(1-\mathcal{A}-2r\mathcal{A}'-\frac{1}{2}r^{2}\mathcal{A}''
\right)
\notag\\
&\qquad {}-\frac{1}{6}(|s|-3|s|^{2}+2|s|^{3})\left(
1-\mathcal{A}+\frac{1}{2}r^{2}\mathcal{A}''
\right)
-\boldsymbol{\lambda}_{(s)}\biggr]R^{(s)}_{l\omega}=G^{(s)}_{l\omega},
\label{radial3}
\end{align}
\textbf{which reduces to \cite{Harris:2003eg} for positive $s$.}

\section{Gauge dependence in gravitational-wave equations}

In previous study two kinds of the gravitational-wave equations have appeared. One is in \cite{Jing:2021ahx, Guo:2023niy},
and the other is in \cite{Jing:2022vks}.
Since those equations are obtained from the same set of the coupled equations in the Newman-Penrose formalism but with the different gauges,
both of the equations should describe the gravitational wave correctly. Here one question can be raised:
although the unknown variables $\Psi_{4}^{B}$ and $\Psi_{0}^{B}$ are the gauge-invariant quantities, why can we have two
(or more in principle) forms of the wave equations for each variable? In order to answer this question,
we will revisit the derivation of the gravitational-wave equation on the background with emphasis of
the gauge dependence.

We here focus on the wave equation for $\Psi_{4}^{B}$ and consider the case of $\mathcal{D}=1$,
because in \cite{Jing:2022vks} this case is only considered. Here our gauge conditions \eqref{gauge01} and \eqref{gauge02}
just reduce to $\lambda^{B}=0$ \cite{Guo:2023niy}.
We will begin with the following three equations in the Newman-Penrose formalism:
\begin{align}
&(\delta+4\beta-\tau)\Psi_{4}-(\Delta+4\mu+2\gamma)\Psi_{3}+3\nu\Psi_{2} \notag\\
&\qquad =(\bar{\delta}-\bar{\tau}+2\bar{\beta}+2\alpha)\Phi_{22}-(\Delta+2\gamma+2\bar{\mu})\Phi_{21}
-2\lambda\Phi_{12}+2\nu\Phi_{11}+\bar{\nu}\Phi_{20}, \label{eq1}\\
&(D+4\epsilon-\rho)\Psi_{4}-(\bar{\delta}+4\pi+2\alpha)\Psi_{3}+3\lambda\Psi_{2} \notag\\
&\qquad =(\bar{\delta}-2\bar{\tau}+2\alpha)\Phi_{21}-(\Delta+2\gamma-2\bar{\gamma}+\bar{\mu})\Phi_{20}
+\bar{\sigma}\Phi_{22}-2\lambda\Phi_{11}+2\nu\Phi_{10}, \label{eq2}\\
&(\Delta+\mu+\bar{\mu}+3\gamma-\bar{\gamma})\lambda-(\bar{\delta}+\pi-\bar{\tau}+\bar{\beta}+3\alpha)\nu+\Psi_{4}=0.
\label{eq3}
\end{align}
We split all the quantities in the above into the background part $(A)$ and the perturbation part $(B)$,
for instance, $\Psi_{4}=\Psi_{4}^{A}+\Psi_{4}^{B}$, etc. We keep the first order of the perturbation only.
The background part of the above equations are indeed satisfied,
and the perturbation part becomes
\begin{align}
&(\delta-4\alpha)^{A}\Psi_{4}^{B}-(\Delta+2\gamma+4\mu)^{A}\Psi_{3}^{B}+3\nu^{B}\Psi_{2}^{A} \notag\\
&\qquad =
\bar{\delta}^{A}\Phi_{22}^{B}-(\Delta+2\gamma+2\mu)^{A}\Phi_{21}^{B}+2\nu^{B}\Phi_{11}^{A},
\label{eq01}\\
&(D-\rho)^{A}\Psi_{4}^{B}-(\bar{\delta}+2\alpha)^{A}\Psi_{3}^{B}+3\lambda^{B}\Psi_{2}^{A}
\notag\\
&\qquad =(\bar{\delta}+2\alpha)^{A}\Phi_{21}^{B}-(\Delta+\mu)^{A}\Phi_{20}^{B}
-2\lambda^{B}\Phi_{11}^{A},
\label{eq02}\\
&(\Delta+2\gamma+2\mu)^{A}\lambda^{B}-(\bar{\delta}+2\alpha)^{A}\nu^{B}+\Psi_{4}^{B}=0.
\label{eq03}
\end{align}
Now we will obtain the wave equation for $\Psi_{4}^{B}$ as the same procedure in the previous section.
We operate $(\Delta+2\gamma+5\mu)^{A}$ to \eqref{eq02} and $(\bar{\delta}+2\alpha)^{A}$ to \eqref{eq01}, respectively,
We then make the difference of them.
The terms with $\Psi_{3}^{B}$ are canceled by \eqref{com1} with $p=2$, $q=-4$
and the remaining becomes
\begin{align}
&\left[(\Delta+2\gamma+5\mu)(D-\rho)-(\bar{\delta}+2\alpha)(\delta-4\alpha)\right]^{A}\Psi_{4}^{B}
\notag\\
&\qquad {}+(3\Psi_{2}+2\Phi_{11})^{A}(\Delta+2\gamma+5\mu)^{A}\lambda^{B}-(3\Psi_{2}-2\Phi_{11})^{A}(\bar{\delta}+2\alpha)^{A}\nu^{B}
\notag\\
&=T_{4}
-\lambda^{B}\Delta^{A}(3\Psi_{2}+2\Phi_{11})^{A}+\nu^{B}\bar{\delta}^{A}(3\Psi_{2}-2\Phi_{11})^{A},
\label{eq04}
\end{align}
where $T_{4}$ is defined by
\begin{align}
T_{4}&=(\Delta+2\gamma+5\mu)^{A}
\left[(\bar{\delta}+2\alpha)^{A}\Phi_{21}^{B}-(\Delta+\mu)^{A}\Phi_{20}^{B}\right]
\notag\\
&\quad {}-(\bar{\delta}+2\alpha)^{A}
\left[\bar{\delta}^{A}\Phi_{22}^{B}-(\Delta+2\gamma+2\mu)^{A}\Phi_{21}^{B}\right].
\end{align}
For the third line in \eqref{eq04}, we have
\begin{align}
\Delta^{A}(3\Psi_{2}-2\Phi_{11})^{A}&=-3\mu^{A}(3\Psi_{2}+2\Phi_{11})^{A}+8\mu^{A}\Phi_{11}^{A},
\label{eqpsi2-1}
\\
\bar{\delta}^{A}(3\Psi_{2}+2\Phi_{11})^{A}&=-3\pi^{A}(3\Psi_{2}-2\Phi_{11})^{A},
\label{eqpsi2-2}
\end{align}
and then substituting the above into \eqref{eq04}, we obtain
\begin{align}
&\left[(\Delta+2\gamma+5\mu)(D-\rho)
-(\bar{\delta}+2\alpha)(\delta-4\alpha)-3\Psi_{2}\right]^{A}\Psi_{4}^{B}
\notag\\
&\qquad {}+2\Phi_{11}^{A}\left[(\Delta+2\gamma+2\mu)^{A}\lambda^{B}
+(\bar{\delta}+2\alpha)^{A}\nu^{B}\right]
\notag\\
&=T_{4}
-4\lambda^{B}[(\Delta+2\mu)\Phi_{11}]^{A},
\label{eq05}
\end{align}
where we have used \eqref{eq03} and $\bar{\delta}^{A}\Phi_{11}^{A}=0$ from the spherical symmetry.
By eliminating the terms with $\nu^{B}$ using \eqref{eq03} again, we have
\begin{align}
&\left[(\Delta+2\gamma+5\mu)(D-\rho)
-(\bar{\delta}+2\alpha)(\delta-4\alpha)-3\Psi_{2}+2\Phi_{11}\right]^{A}\Psi_{4}^{B}
\notag\\
&=T_{4}
-4(\Delta+2\gamma+4\mu)^{A}(\Phi_{11}^{A}\lambda^{B}).
\label{eq06}
\end{align}
Moreover, using \eqref{eqpsi2-1} and \eqref{eqpsi2-2} again, the above can be rewritten as
\begin{align}
&\bigl[(\Delta+2\gamma+5\mu)(D-\rho)
-(\bar{\delta}+2\alpha)(\delta-4\alpha)-3\Psi_{2}+2\Phi_{11}\bigr]^{A}
\Psi_{4}^{B}
\notag\\
&=T_{4}+(3\Psi_{2}-2\Phi_{11})^{A}(\Delta+2\gamma+2\mu)^{A}\lambda^{B}
-(\Delta+2\gamma+5\mu)^{A}\left[(3\Psi_{2}+2\Phi_{11})^{A}\lambda^{B}\right],
\label{eqpsi4-2}
\end{align}
So far we have not used any gauge condition. If we take the gauge condition as $\lambda^{B}=0$,
then the equations \eqref{eq05} and \eqref{eqpsi4-2} is reduced to the wave equation \eqref{GW04} in \cite{Jing:2021ahx, Guo:2023niy} as
\begin{align}
\bigl[(\Delta+2\gamma+5\mu)(D-\rho)
-(\bar{\delta}+2\alpha)(\delta-4\alpha)-3\Psi_{2}+2\Phi_{11}\bigr]^{A}
\Psi_{4}^{B}
=T_{4},
\label{GW1}
\end{align}
which has a similar form as the vacuum case $\Phi_{11}^{A}=\Lambda^{A}=0$.
On the other hand, from \eqref{eq02}, $\lambda^{B}$ can be expressed in terms of $\Psi_{3}^{B}$ as
\begin{gather}
\lambda^{B}=\frac{1}{(3\Psi_{2}+2\Phi_{11})^{A}}
\left[-(D-\rho)^{A}\Psi_{4}^{B}+(\bar{\delta}+2\alpha)^{A}(\Psi_{3}+\Phi_{21})^{B}-(\Delta+\mu)^{A}\Phi_{20}^{B}\right].
\end{gather}
Substituting the above into \eqref{eqpsi4-2} gives
\begin{align}
&\left[F_{2}^{-1}(\Delta+2\gamma+2\mu-F_{1})(D-\rho)-(\bar{\delta}+2\alpha)(\delta-4\alpha)-3\Psi_{2}+2\Phi_{11}\right]^{A}\Psi_{4}^{B}
\notag\\
&=F_{2}^{-1}(\Delta+2\gamma+2\mu-F_{1})^{A}\left[(\bar{\delta}+2\alpha)^{A}(\Psi_{3}+\Phi_{21})^{B}-(\Delta+\mu)^{A}\Phi_{20}^{B}\right]
\notag\\
&\qquad{}-(\bar{\delta}+2\alpha)^{A}\left[\bar{\delta}^{A}\Phi_{22}^{B}-(\Delta+2\gamma+2\mu)^{A}\Phi_{21}^{B}\right],
\label{eqpsi4-3}
\end{align}
or
\begin{align}
&\left[(\Delta+2\gamma+2\mu-F_{1})(D-\rho)-F_{2}(\bar{\delta}+2\alpha)(\delta-4\alpha)-3\Psi_{2}-2\Phi_{11}\right]^{A}\Psi_{4}^{B}
\notag\\
&=(\Delta+2\gamma+2\mu-F_{1})^{A}\left[(\bar{\delta}+2\alpha)^{A}(\Psi_{3}+\Phi_{21})^{B}-(\Delta+\mu)^{A}\Phi_{20}^{B}\right]
\notag\\
&\qquad{}-F_{2}(\bar{\delta}+2\alpha)^{A}\left[\bar{\delta}^{A}\Phi_{22}^{B}-(\Delta+2\gamma+2\mu)^{A}\Phi_{21}^{B}\right],
\label{eqpsi4-4}
\end{align}
where $F_{1}$ and $F_{2}$ are defined by
\begin{gather}
F_{1}=\Delta\left[\ln(3\Psi_{2}+2\Phi_{11})^{A}\right], \quad F_{2}=\left(\frac{3\Psi_{2}+2\Phi_{11}}{3\Psi_{2}-2\Phi_{11}}\right)^{A}.
\end{gather}
Thus if we take the gauge condition as $\Psi_{3}^{B}=0$, then \eqref{eqpsi4-4} is reduced to the wave equation in \cite{Jing:2022vks} as
\begin{gather}
\left[(\Delta+2\gamma+2\mu-F_{1})(D-\rho)-F_{2}(\bar{\delta}+2\alpha)(\delta-4\alpha)-3\Psi_{2}-2\Phi_{11}\right]^{A}\Psi_{4}^{B}
=\tilde{T}_{4},
\label{GW2}
\end{gather}
where $\tilde{T}_{4}$ is defined by
\begin{align}
\tilde{T}_{4}
&=(\Delta+2\gamma+2\mu-F_{1})^{A}\left[(\bar{\delta}+2\alpha)^{A}\Phi_{21}^{B}-(\Delta+\mu)^{A}\Phi_{20}^{B}\right]
\notag\\
&\qquad{}-F_{2}(\bar{\delta}+2\alpha)^{A}\left[\bar{\delta}^{A}\Phi_{22}^{B}-(\Delta+2\gamma+2\mu)^{A}\Phi_{21}^{B}\right].
\label{tilT}
\end{align}
Next we will consider the gauge transformation in the wave equation, as which we take the following tetrad rotations%
\footnote{Since we here consider the case $\mathcal{D}=1$ only, the tetrad rotations are enough.
For the genetral case $\mathcal{D}\neq 1$, we have to take into account the general coordinate transformation as well \cite{Guo:2023niy}. }
\cite{Janis:1965tx}:
\begin{align}
& l^{\mu}\to l^{\mu},\quad m^{\mu}\to m^{\mu}+al^{\mu}, \quad \bar{m}^{\mu}\to \bar{m}^{\mu}+\bar{a}l^{\mu},
\quad n^{\mu}\to n^{\mu}+\bar{a}m^{\mu}+a\bar{m}^{\mu}+a\bar{a}l^{\mu},
\label{rot01}\\
& n^{\mu}\to n^{\mu},\quad m^{\mu}\to m^{\mu}+bn^{\mu}, \quad \bar{m}^{\mu}\to \bar{m}^{\mu}+\bar{b}n^{\mu},
\quad l^{\mu}\to l^{\mu}+\bar{b}m^{\mu}+b\bar{m}^{\mu}+b\bar{b}n^{\mu},
\label{rot02}\\
& l^{\mu}\to e^{-c}l^{\mu},\quad n^{\mu}\to e^{c}n^{\mu},\quad
m^{\mu}\to e^{i\vartheta}m^{\mu},\quad \bar{m}^{\mu}\to e^{-i\vartheta}\bar{m}^{\mu}.
\label{rot03}
\end{align}
In order not to change the background, we assume that the parameters $a$, $b$, $c$ and $\vartheta$ are in the first order of
the perturbation, and hence for the perturbation quantities, the transformation of the first order is enough.
In the wave equation \eqref{eq05}, the left hand side is gauge-invariant since $\Psi_{4}^{B}$ is so, which imples that
the right hand side has to be invariant as well.
Using
\begin{gather}
\lambda^{B}\to\lambda^{B}+(\bar{\delta}+2\alpha)^{A}\bar{a},
\quad \Phi_{21}^{B}\to\Phi_{21}^{B}+2\Phi_{11}^{A}\bar{a},
\quad \Phi_{20}^{B}\to\Phi_{20}^{B}, \quad \Phi_{22}^{B}\to\Phi_{22}^{B},
\end{gather}
The source term $T_{4}$ transforms as
\begin{gather}
T_{4}\to T_{4}+4\left[(\bar{\delta}+2\alpha)(\Delta+2\gamma+3\mu)\right]^{A}(\Phi_{11}^{A}\bar{a}).
\end{gather}
We can find that this transformation is cancelled by that of other terms in the right hand side of \eqref{eq05}, where
we have used the formula \eqref{com1} and $\bar\delta^{A}\Phi_{11}^{A}=0$.
We can also show that $\tilde{T}_{4}$ defined by \eqref{tilT} has nontrivial gauge transformation under \eqref{rot01}--\eqref{rot03},
which is cancelled by that of $\Psi_{3}^{B}$ as
\begin{gather}
\Psi_{3}^{B}\to\Psi_{3}^{B}+3\Psi_{2}^{A}\bar{a}.
\end{gather}
Thus the origin of the gauge dependence of the gravitational-wave equation is due to that of the source term, in particular $\Phi_{21}^{B}$.
We note that in the vacuum case this dependence does not appear since $\Phi_{11}^{A}=0$.
We can also show that two gravitational-wave equations \eqref{GW1} and \eqref{GW2} coincide in the vacuum background because of
\begin{gather}
F_{1}=-3\mu^{A}, \quad F_{2}=1,
\end{gather}
for the vacuum.

\section{Summary and discussion}

In this paper we have studied the wave equations with the various spins on the background of the general spherically symmetric spacetime.
By introducing the spin variable $s$, we have unified those equations using $s$ itself, $|s|$ and $\delta_{s}$.
The transformation \eqref{redef1} to make the equation simplified is possible, and the form of \eqref{redef1} in turn becomes the same
as the vacuum case, although the background, in particular the spin coefficient $\rho^{A}$, is deformed.

We have also discussed the gauge dependence in the form of the gravitational-wave equations in the previous study.
The gauge dependence of the wave equation is in turn originated from that of the source term,
and hence it seems one cannot avoid it, unless the vacuum case. If we take another gauge, the form of the gravitational-wave equation will be changed,
which will also affect to the unified expression \eqref{unified1} and \eqref{unified2}.

A similar analysis with this paper using the metric perturbation has been performed in \cite{Thompson:2016fxe, Lenzi:2021wpc, Liu:2022csl},
where the wave equations are  resemble to the Regge-Wheeler and the Zerilli equations \cite{Regge:1957td,Zerilli:1970se},
of which the generalization to the general spin $s$ has also been studied \cite{Leaver:1986gd}.
In the vacuum case, there is a transformation between the Teukolsky and the Regge-wheeler equation, which is called
the Chandrasekhar transformation \cite{Chandrasekhar:1975}. Moreover, for $s \neq 0$,  the relation between the $R^{(-|s|)}_{l\omega}(r)$
and $R^{(|s|)}_{l\omega}(r)$ can also be regarded as the special case of the Chandrasekhar transformation \cite{Teukolsky:1973ha,Starobinsky1}.
It would be interesting to see whether similar relations hold in the current case or not \textbf{as in \cite{Arbey:2021yke}}.

A possible generalization would be to extend the results here to the axisymmetric background, which contains the Kerr black hole as an example.
In the vacuum case the backgrounds satisfying the type D condition are fully classified, so-called the Kinnersley metric \cite{Kinnersley:1969zza}.
It would be interesting to find the non-vacuum extension of the Kinnersley metric and the wave equation on that background with the general spin.
The gravitational-wave equation on a certain non-vacuum axisymmetric background has been proposed in \cite{Jing:2023vzq}.
However, that equation does not allow the separation of the variables, and the further modification is then needed \cite{Guo:2023wtx}.

\section*{Acknowledgements}
The authors thank Prof. Remo Ruffini for useful discussions. This work was supported in part by the National Natural Science Foundation of China (Grant No. 11973025).



\end{document}